\documentstyle[pre,twocolumn,aps,epsfig]{revtex} 
\begin{document}
\draft
\title{Cycle expansions for intermittent diffusion}
\author{C. P. Dettmann and Predrag Cvitanovi\'{c}}
\address{Dept. of Physics and Astronomy, Northwestern University, Evanston Il
60208, USA \\
and\\
Center for Chaos and Turbulence Studies, Niels Bohr Institute \\
Blegdamsvej 17, DK-2100 Copenhagen {\O}, Denmark\\
dettmann@nbi.dk, p-cvitanovic@nwu.edu}
\date{\today}
\maketitle
\begin{abstract}
We investigate intermittent diffusion using cycle expansions, and show that
a truncation based on cycle stability achieves reasonable convergence.
\end{abstract}
\pacs{PACS: 5.45.+b}

\section{Introduction}

Classical dynamical systems range from purely integrable to purely
hyperbolic.  For purely integrable systems we have a variety of
classical methods, such as separation of the Hamilton-Jacobi
equation~\cite{Go}. 
For almost integrable systems we have KAM
theory~\cite{BGGS}.  For purely hyperbolic systems it is possible
to obtain much information about the system by 
grouping contributions computed on unstable periodic orbits~\cite{ruelle} 
into terms in cycle expansions~\cite{AAC,QCcourse}.
They yield the classical escape rate
of open billiard systems to a high degree of accuracy~\cite{CE}, 
and 
the semiclassical 
energy levels of systems such as helium~\cite{WRT} using
a surprisingly small number of unstable periodc orbits.

However, the formalism does not work well for generic dynamical flows
for which the hyperbolic regions coexist with attractors,
intermittent regions and elliptic regions.
For intermittent systems
the cycle expansions ordered by the topological cycle length
converge poorly if
very long almost stable cycles dominate the dynamics.
The original~\cite{AAC} as well as more
recent applications~\cite{OpusDahlqvistrum} of cycle expansions to intermittent systems 
use detailed analytic information about the intermittent regions
in order to explicitly sum infinite sequences of 
such cycles. 

Our philosophy here is that 
it should be possible to obtain reliable
dynamical averages without a complete understanding of the detailed
structure of the phase space, as long as we are restricted to a given
connected region in which the dynamics is ergodic.  
Recent work~\cite{DM97} on the Lorentz gas
suggests that reordering the cycle expansions 
by stability~\cite{DR91}
may improve convergence in such situations.  Here we test this
proposal by calculating
diffusion in a one-dimensional intermittent map and demonstrate
that the stability ordering yields better convergence than the
ordering by the topological cycle length.

There are several arguments in favour of using stability rather than the
topological or (in case of continuous flows) real time length
as the truncation criterion:
\begin{enumerate}
\item Longer but less unstable cycles can give larger contributions
to a cycle expansion than short but highly unstable cycles.
In such situation truncation by length may require
an exponentially large number of very unstable cycles before a significant
longer cycle is first included in the expansion.

\item Stability truncation requires only that all cycles up to given
stability cutoff be determined, without 
requiring detailed understanding of the topology of the
flow and symbolic dynamics. It is thus much easier to implement for
a generic dynamical system than the
curvature expansions~\cite{AAC} which rely on finite subshift approximations
to a given flow.

\item 
The stability ordering
preserves approximately any shadowing that is present.  That is, a long
cycle which is shadowed by several shorter ones will have a stability
eigenvalue which is approximately the product of the shorter cycle
eigenvalues, and will be most likely be included at the same 
stability cutoff.

\item Cycles can be detected numerically by
searching a long trajectory for near recurrences~\cite{auerbach,MR}.  
The method preferentially finds 
the least unstable cycles, regardless of their topological length.
Another practical advantage of the method (in contrast to
the Newton method searches)
is that it only finds
cycles in a given connected ergodic component of phase space, 
even if isolated cycles or other ergodic regions
exist elsewhere in the phase space. 
\end{enumerate}

In what follows we illustrate the first three points by investigating
the convergence of stability cutoff approach
for a simple system.  We begin by describing
diffusion on a lattice of one-dimensional maps, how to calculate the
diffusion coefficient using cycle expansions, and then perform the
calculations numerically.  Finally we discuss the scope of such
approaches and possible improvements.

\section{Diffusion in 1-D maps}

As a model on which to test the above ideas we shall use a
well understood one parameter family of diffusive one-dimensional
maps.  For such maps the symbolic dynamics is a complete binary shift,
all cycles can be exhaustively enumerated, and the limitations of
the length truncated cycle expansions are solely due to the lack of 
hyperbolicity, and not to inadequate understanding of the symbolic 
dynamics.

Many of the features of intermittent systems can be captured by 
one-dimensional intermittent maps introduced in ref.~\cite{PM} 
to study turbulence.  Piecewise linear approximations~\cite{GW}
can make statistical mechanics aspects of
such intermittent dynamics, including phase transitions~\cite{phtrans}
and Levy flights, analytically tractable.
Intermittent maps can lead to
anomalous deterministic diffusion, with the mean square displacement either
sublinear or superlinear in the time~\cite{GT}.  

In the interval $\hat{x}\in[-1/2,1/2)$, which we call the elementary cell,
our model map takes the form
\begin{equation}
\hat{f}(\hat{x})=\hat{x}(1+2|2\hat{x}|^\alpha)\;\;,
\label{e:map}
\end{equation}
where $\alpha>-1$.  For any value of $\alpha$, this maps the interval
monotonically to $[-3/2,3/2)$.  Outside the elementary cell, the map is
defined to have a discrete translational symmetry,
\[ 
\hat{f}(\hat{x}+n)=\hat{f}(\hat{x})+n\qquad n\in\cal{Z}\;\;.
\] 
A typical initial $\hat{x}$ in the elementary cell diffuses,
wandering over the real line.  The map is parity symmetric,
$
\hat{f}(-\hat{x})=-\hat{f}(\hat{x})\;\;,
$
so the average value of $\hat{x}_{n+1}-\hat{x}_n$ is zero,
and there is no mean drift.

\begin{figure}
\vspace*{8cm}
\includegraphics{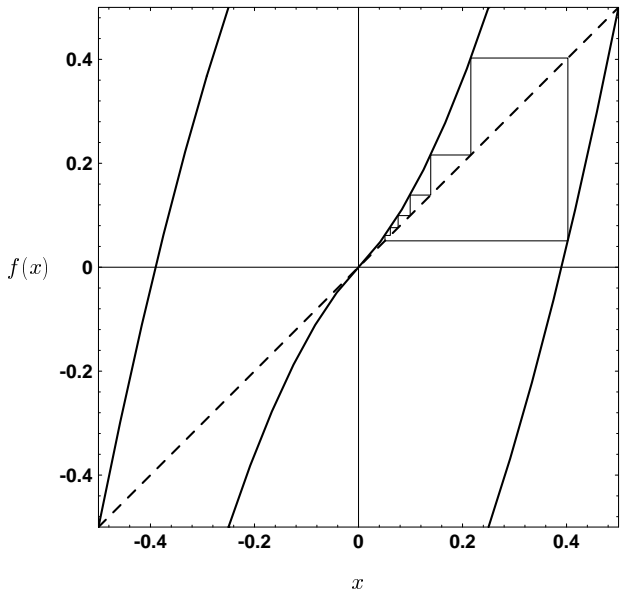}
\caption[]{
The map (\ref{e:map}) for $\alpha=1$
restricted to the elementary cell, together with the
$0^6+$ cycle. In the full space this cycle
correspond to ballistic motion to the right, with each 7th iteration
entering the next cell.
}
\label{f:map}
\end{figure}

We now restrict the dynamics to the elementary cell, that is, we define
\[
x=\hat{x}-[\hat{x}+1/2]\;\;,
\]
where $[z]$ is the greatest integer less than or equal to $z$,
so that   $x$ is restricted to the range $[-1/2,1/2)$.  
The reduced map (see Fig.~\ref{f:map}) is
\begin{equation}
f(x)=\hat{f}(x)-[\hat{f}(x)+1/2]\;\;.
\label{e:RedMap}
\end{equation}

A cycle $p=\{x_1,x_2,\ldots,x_n\}, f^{n}(x_j)=x_j$ with stability
$\Lambda_p=\prod_{j=1}^{n}f'(x_j)$ corresponds to a
trajectory which returns to an equivalent point in the full dynamics,
$\hat{f}^n(x_j)=x_j+\hat{n}_p$.  Thus the cycles fit into two categories,
those which are periodic in the full dynamics, $\hat{n}_p=0$, and those
which are not.  The diffusive properties of the map are fully specified
by the reduced map $f(x)$, together with the lattice translation
$\hat{n}$.  The diffusion constant
\begin{equation}
D=\lim_{n\rightarrow\infty}\frac{1}{2n}\langle \hat{n}^2 \rangle
\label{e:ein}
\end{equation}
is computed as an average over initial conditions in the elementary
cell. 

The reduced map (\ref{e:RedMap})
has three branches, corresponding to either moving to the left,
staying in the elementary cell, or moving to the right, 
$\hat{n}=\{-1,0,1\}$.
Hence the natural symbolic dynamics is a 3-letter alphabet $\{-,0,+\}$.
For a given symbol string the total translation 
$\hat{n}$ is just the sum of the individual symbols in the cycle
symbol string.  The three branches form a Markov partition,
because each is mapped onto
the whole interval, and the symbolic dynamics is thus unrestricted in the three
symbols, with all finite strings corresponding to cycles.  

The point $x=0$ is a fixed point (cycle of length 1) with symbol
sequence $0$.  
For
$\alpha<0$, $\Lambda_0=f'(0)=\infty$, this fixed point
is infinitely unstable, and its contribution
to cycle expansions is vanishing.  For $\alpha>0$, $\Lambda_0=f'(0)=1$,
the fixed point is marginally stable and is also customarily omitted from
cycle expansions~\cite{AAC}.
The intermittent behavior arises from cycles containing long strings of $0$'s
which come close to the marginally stable fixed point.

\section{Cycle expansions}

Cycle expansion
approaches to deterministic diffusion in one-dimensional maps were
introduced in refs.~\cite{art91,ACL1} and in the Lorentz gas in
ref.~\cite{Va}.  
The dynamical zeta function formula~\cite{QCcourse}
for the diffusion coefficient is
\begin{equation}
D=\frac{1}{2}\frac{\sum (-1)^k 
   {(\hat{n}_{p_1}+\hat{n}_{p_2}+\ldots+\hat{n}_{p_k})^2
     \over |\Lambda_{p_1}\Lambda_{p_2}\ldots\Lambda_{p_k}|}
                  }{
                   \sum (-1)^k 
   {n_{p_1}+n_{p_2}+\ldots+n_{p_k}
     \over |\Lambda_{p_1}\Lambda_{p_2}\ldots\Lambda_{p_k}|}
                   }
\;\;,\label{e:cyc}
\end{equation}
where the sum is over nonempty distinct nonrepeating combinations of
prime cycles, $\hat{n}$ is the lattice translation of a cycle,
$n$ is the period of the cycle,  
and $\Lambda$ its stability.
As the flow is conserved, the leading eigenvalue of the
Frobenius-Perron operator equals unity, and the inverse of the
corresponding dynamical zeta function~\cite{ruelle}
must vanish~\cite{QCcourse} for $z=1$:
\begin{equation}
1/\zeta(1)=1+\sum (-1)^k/|\Lambda_{p_1}\Lambda_{p_2}\ldots\Lambda_{p_k}|=0
\;\;.
\label{zet}
\end{equation}
For example, with the $\{-,0,+\}$ symbolic dynamics 
the cycle expansion up to topological length $n=2$ equals
\[
1/\zeta(1)=1-{1\over \Lambda_{+}} -{1\over \Lambda_{-}} 
	     -{1\over \Lambda_{+0}} - {1\over \Lambda_{-0}} 
      -{1\over \Lambda_{+-}} +{1\over \Lambda_{+}\Lambda_{-}}\;\;,
\]
where we have omitted the $0$ cycle.

Ideally the $+-$ cycle is shadowed by the 
$+$ and $-$ cycles,
so the last two terms are expected to approximately cancel.  The
cancellation is exact
when $\alpha=0$ and both terms equal $9$. However, for other 
values of $\alpha$ 
shadowing may not lead to any significant cancellations.
For example, for $\alpha=1$, 
$\Lambda_{+}\Lambda_{-}=25$ but $\Lambda_{+-}\approx 12$.

For the case $\alpha=0$ the map $f(x)$ is piecewise linear,
$\Lambda=3^n$ for all cycles, and in this case the stability and
length ordering are equivalent.  If the $0$ cycle is included,
all $n>1$ terms in (\ref{e:cyc}) cancel, leading to $D=1/3$.  

For $\alpha>0$, the dynamics is intermittent, 
and $\Lambda_p$ does not necessarily grow exponentially with
cycle length $n_p$.
Furthermore, for $\alpha>0$ the most stable orbits
which contain long strings of $0$'s are not shadowed by
combinations of shorter cycles, as the $0$ cycle is not included.
In fact, we can explicitly deduce the behavior of the most stable
cycles, those of the form $0^n+$
(an example is given in Fig.~\ref{f:map}).  These begin at initial point
which we shall denote by $x_n$, 
slightly greater than zero.  Many iterations of the function increase
monotonically the value of $f^k(x)$ until it finally crosses over to
the right branch and returns
to the starting point.  Inserting an extra $0$ in the symbolic
dynamics has the effect of slightly decreasing the starting point, but
the other cycle values are virtually unchanged, 
the new starting point is very close to the old one, and
$f(x_{n+1})=x_n$ is a good approximation for moderately large $n$.
Thus for $\alpha>0$
\[ 
\frac{x_{n}}{x_{n+1}} \approx 1+2(2x_{n+1})^{\alpha}
\] 
to the leading order in $x_n$.
This difference equation may be approximately solved as a power law,
$x_n=\gamma n^{-\delta}$ giving $\alpha\delta=1$,
$2(2\gamma)^{\alpha}=\delta$, or
\[
x_n=\frac{1}{2 (2\alpha n)^{1/\alpha} }\;\;.
\]
The stability can be estimated in a similar fashion:
\[ 
\frac{\Lambda_{n}}{\Lambda_{n-1}}
	\approx
		f'(x_n)=1+2(\alpha+1)(2x_n)^{\alpha}
=1+\frac{\alpha+1}{\alpha n}\;\;.
\] 
Again putting $\Lambda_n=\rho n^{\epsilon}$ we obtain $\epsilon=1+1/\alpha$,
\[
\Lambda_n\sim n^{1+1/\alpha}\;\;,
\]
confirmed by our numerical results.  This power-law growth of 
$\Lambda_n$ is in contrast
to hyperbolic systems for which all cycles have stabilities which 
grow exponentially with the cycle length.

Because these are the most stable cycles, they dominate cycle expansions
at given $n$.  Combinations of cycles which do not include a cycle with
a string of almost $n$ $0$'s are highly suppressed. For example, two
cycles with $n/2$ $0$'s and one other symbol each have a combined stability
\[ 
\Lambda_{n/2}^2\sim (n^2/4)^{1+1/\alpha}\gg \Lambda_n
\] 
for large $n$, again in contrast with hyperbolic systems, for which 
such shadowing combinations are of comparable magnitude.  
Hence we can estimate the convergence properties of 
such cycle expansions
by approximating them with the dominant $0^n+$ 
cycle family~\cite{AAC}.
In this approximation the flow conservation condition (\ref{zet})
\begin{equation}
1/\zeta(1)\sim\sum_n {1 \over n^{1+1/\alpha}}\;\;,
\label{e:ApprxNormCond}
\end{equation}
is approximately the Riemann zeta function $\zeta(1+1/\alpha)$,
convergent for all $\alpha>0$, which is just as well.  The denominator
in the diffusion formula (\ref{e:cyc})
appears whenever we calculate
the time average of some quantity, and plays the role of
a mean cycle period
\begin{equation}
\langle n\rangle_{\zeta}\sim\sum_n {1 \over n^{1/\alpha} }\;\;.
\label{e:MeanCycPer}
\end{equation}
As $\alpha$ increases, the system spends more and more of its time near
the marginally stable fixed point, and
for $\alpha\geq 1$, the system
spends on average all of its time within an arbitrarily small neighborhood
of this point, leading to a divergent mean cycle period (\ref{e:MeanCycPer}).
The numerator of the expression for diffusion looks like the
flow conservation sum (\ref{e:ApprxNormCond}), 
with extra factors of $\hat{n}_p^2$.  This factor is
a number of order unity for the least unstable cycles, so the series
converges.
Thus the average (\ref{e:cyc}) which defines the
diffusion coefficient undergoes a phase transition~\cite{phtrans}
and equals zero for $\alpha\geq 1$.  This behavior is described
as ``weak'' ($0<\alpha<1$) or ``strong'' ($\alpha\geq 1$) intermittency.
Other averages may converge for different ranges of $\alpha$.

For $\alpha\geq 1$ the diffusion is anomalous, with $\langle \hat{n}^2 \rangle$
increasing more slowly than $n$.  
In the case at hand
\[ 
1/\zeta(z)\sim\sum_n {z^n \over n^{1+1/\alpha} }
\sim\left\{\begin{array}{ll}
(z-1)&\alpha<1\\
(z-1)\ln(z-1)&\alpha=1\\
(z-1)^{1/\alpha}&\alpha>1
\end{array}\right.\;\;,
\] 
and the leading behavior of a dynamical zeta
function as $z\rightarrow 1$  yields~\cite{ACL1}
the exponent characterizing the sublinear diffusion
\[ 
\langle x_n^{2} \rangle\sim\left\{\begin{array}{ll}
n&\alpha<1\\
n/\ln n&\alpha=1\\
n^{1/\alpha}&\alpha>1
\end{array}\right.\;\;.
\] 

We are now in position to estimate and compare
the rates of convergence of the topological and stability truncation approaches.
As we have seen, the cycle expansions
are dominated by terms of the form $1/n^{\gamma}$ where $\gamma=1+1/\alpha$
for the flow conservation condition, and $\gamma=1/\alpha$ for the diffusion constant.
Thus the error made by the topological
truncation length after $n$ terms is 
$\sim n^{1-\gamma}$.  In contrast,  truncating by stability
corresponds to an error of order of
$\Lambda^{-1/(\alpha+1)}$ for the flow conservation, and
$\Lambda^{(\alpha-1)/(\alpha+1)}$ for the diffusion constant.  For
$\alpha \to 1$ close to where
the expansion diverges it may
be advisable to improve the estimates by convergence acceleration techniques.

Having defined the cycle expansions and analyzed
their behavior in the intermittent case, we now pose the more pragmatic
question: What is the optimal ordering
in practice?  In the approximation we have been using,
with only a single family of cycles contributing, the ordering is 
self-evident. However, the full expansion is only conditionally
convergent, and as we have no proof that the stability
ordering yields the correct results, our justification will come from
heuristic arguments, together with the numerical results.  

The topological length cutoff corresponds to a complete
partitioning of the phase space into $3^{n}$ periodic point
neighborhoods, irrespective of the relative sizes of these neighborhoods.
What is the meaning of the stability cutoff? A cycle expansion
can be interpreted~\cite{QCcourse} as a partition of the dynamical
phase space into neighborhoods of periodic points $i\in p$, each of
size $\sim 1/\Lambda_p$. A fixed stability cutoff $\Lambda$ selects a
uniform partition of the phase space into $\sim \Lambda$ regions, each
of period approximately $N = \ln \Lambda$, the time needed for a neighborhood
of a hyperbolic orbit to spread across the entire system. Each prime
cycle has about $N$ periodic points, so
the number of prime cycles up to a given
stability grows as  $\Lambda/\ln\Lambda$.
This estimate is known as the dynamics version of the prime number theorem
for Axiom A systems,  given in ref.~\cite{PP}. 
We find that the estimate is valid numerically for nonhyperbolic systems
as well, in the case at hand for all values of $\alpha> -1$.

The dramatic difference between the two approaches
is the number of cycles required in each case.  The
number of prime cycles up to a given length increases exponentially with the
length, in our case as $\sim 3^n/n$.  
The number of prime cycles up to a given
stability is grows as  $\Lambda/\ln\Lambda$.
Superficially,
the topological ordering requires an exponential number of cycles,
while the stability ordering requires only a power law.
The issue is how small is the error
for a given truncation. In the case of nice hyperbolic flows 
this error is superexponentially small, but for intermittent
systems, the size of the error is not known, and we have to
resort to numerics to estimate it.

\section{Numerical results}

\begin{figure}
\vspace*{14.5cm}
\includegraphics{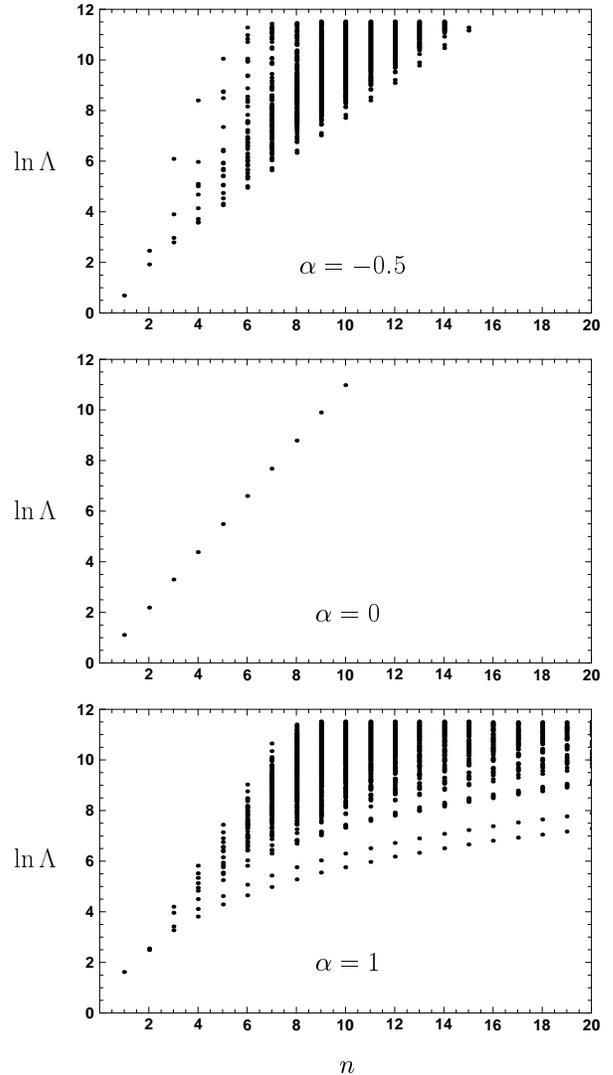}
\caption[]{
The distribution of cycle eigenvalues as a function of
the topological cycle length:
For $\alpha=1$, the intermittent case,
a fixed topological length $n$ cutoff misses
many of the least unstable (but longer period) cycles.
}
\label{f:CycStabs}
\end{figure}

For a simple one-dimensional map with a complete symbolic dynamics,
such as the map (\ref{e:map}) studied here, 
almost any reasonable cycle finding method~\cite{QCcourse}
should yield thousands of prime cycles.
The accuracy of $1/\Lambda$'s that we calculate
approaches the machine 
precision.  
We implement the stability ordering by noting that
for this map any cycle
containing an extra symbol is  
less stable than the preceding one. 
We recursively increment cycle lengths, starting with $+$ and $-$, and
stopping when the stability cutoff is reached.
The stability ordering is fast, as cycles containing large numbers of
$0$'s which dominate the expansion appear only a few times in this
enumeration.
The distribution
of cycles as a function of $n_p$ and $\Lambda_p$ is shown in 
Fig.~\ref{f:CycStabs}.

\begin{figure}
\vspace*{4.5cm}
\includegraphics{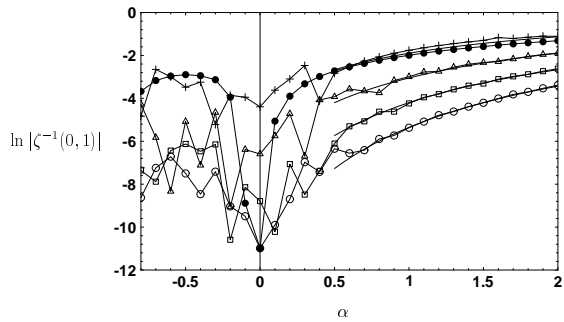}
\caption[]{
Numerical test of the flow conservation for different truncations.
The logarithm of the magnitude of the zeta function (\ref{zet})
is plotted as a function of $\alpha$
for stability cutoffs $\Lambda=$
$10^2$, $10^3$, $10^4$, and $10^5$ (plusses, open triangles, squares
and circles, respectively), as well as for the topological length
cutoff $N=10$ (filled circles).  The solid curves for $\alpha>0.5$
are the behavior expected from the previous section,
$\ln\left(1.5\Lambda^{-\alpha-1}\right)$, for the cutoff values of
$\Lambda$, with the constant $1.5$ chosen to fit the data.
}
\label{f:norm}
\end{figure}

In Fig.~\ref{f:norm} we present a comparison of
how well different truncations 
respect the flow conservation rule (\ref{zet}).
There is a definite improvement as the stability cutoff
is increased, demonstrating convergence.  For smaller values of
$\alpha$ there is a significant amount of scatter due to a small
number of unbalanced shadowing terms which vary rapidly with $\alpha$,
however the error is consistently small. For example,  for $\alpha<1$
the error curve
corresponding to $\Lambda=10^5$ cutoff always lies below 
$e^{-6} \approx 0.0025$ .

The filled circles in Fig.~\ref{f:norm} are obtained by using all cycles
with $N \leq 10$, corresponding to roughly the same computational effort
as the stability cutoff of $10^5$.  The error is smooth but large,
comparable to the stability cutoff only at $\alpha=0$, the solvable
case.

\begin{figure}
\vspace*{5.5cm}
\includegraphics{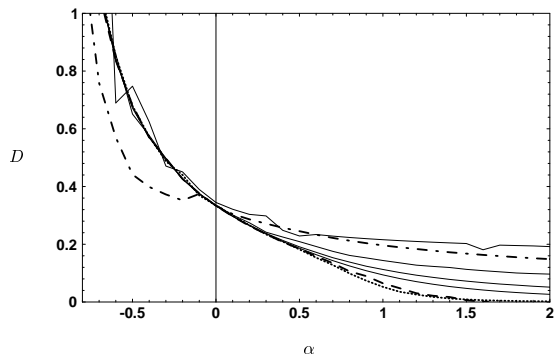}
\caption[]{
Diffusion constant (\ref{e:ein})
as function of $\alpha$ computed by 3 different methods.
The dotted line is obtained by direct
numerical evaluation of (\ref{e:ein}) for $3\times 10^4$ random initial
points, each evolved by $3\times 10^4$ iterations of the map.
The dot-dash line is 
the cycle expansion average (\ref{e:cyc})
truncated at topological length $N=10$, corresponding to $9381$ cycles
with $0$ omitted.
The expansion is accurate only near $\alpha=0$,
where the shadowing is exact.
The solid lines are cycle expansions (\ref{e:cyc})
truncated according to stability, with cutoff
$\Lambda=10^2$ for the uppermost line at $\alpha=2$,
followed by $10^3$, $10^4$ and $10^5$ for the other
curves.  The dashed line is an extrapolation~(\ref{e:d2}) of these
curves.
 $\alpha=1$ corresponds to
a phase transition point beyond which $D=0$; all numerical methods fail
here.
}
\label{f:diff}
\end{figure}

The diffusion constant, evaluated using three different methods, is
plotted as a function of $\alpha$ in Fig.~\ref{f:diff}.  Each of the
three methods used about half an hour of computing time for each
value of $\alpha$.  There is no analytic expression for $D$ in
general, except for $D=1/3$ at $\alpha=0$ and $D=0$ for
$\alpha\geq 1$, as explained above.

The cycle expansions (\ref{e:cyc}) truncated at topological length
$N=10$ (dot-dash line) give very
poor estimates of $D$ as $\alpha$ increases away from zero into the
intermittent regime, as many of the least unstable cycles are not
included (Fig.~\ref{f:CycStabs}).
More surprisingly, 
even for $\alpha<0$, the hyperbolic phase dominated by exponentially many
unstable cycles, the estimate is poor, this time
in the opposite direction, because there are many omitted cycles near
the $+$ and $-$ 
cycles with large displacements and moderate stability,
$\Lambda\sim\Lambda_+^n$.  The stability of the 
$+$ and $-$ cycles is
$\Lambda_+=3+2\alpha$ which is relatively small in this case.

The cycle expansions (\ref{e:cyc}) truncated according to stability
(solid lines) approach the direct simulation (dotted line) as the
cutoff is increased from $10^2$ (upper curve at $\alpha=2$) to $10^5$.
Since the number of prime cycles less than $\Lambda$ is
asymptotically equal to $\Lambda/\ln\Lambda$, we expect that about
$10^5/\ln 10^5 \approx 8700$ cycles would be needed, not 
far from the number of
cycles actually found, which lies in the range $8871$-$10066$
for all the $\alpha$ shown.

Due to the phase transition
all three methods are only logarithmically convergent to zero at $\alpha=1$. 
Direct simulation cannot
yield accurate estimates of $D$ near this point; the stability and
topological length truncations could easily be improved
by using the analytic structure of the $0^n+$
series~\cite{OpusDahlqvistrum} for the map at hand, 
but as we are unlikely to have this information available in a generic case,
such improvements are outside the scope of the present investigation.

As discussed in the previous section, we expect that the error should
scale as a power of the cutoff stability, related to $\alpha$, or
exponentially with $n$ for a cutoff of $10^n$.  If the convergence of
a sequence is exponential and reasonably smooth, it should be possible
to extrapolate to the limit using a variety of convergence 
acceleration schemes. We have tried Aitken's $\delta^2$-process~\cite{PTVF}.
If $t_{n-1}$, $t_n$ and $t_{n+1}$ are three consecutive terms of a
sequence, an improved estimate is
\begin{equation}
t_n^{'}=t_{n+1}-\frac{(t_{n+1}-t_n)^2}{t_{n+1}-2t_n+t_{n-1}}\;\;.
\label{e:d2}
\end{equation}
Using this formula for the cycle expansions computed to a stability
of $10^3$, $10^4$ and $10^5$ we obtain the dashed line in Fig.~\ref{f:diff},
which is comparable to the direct simulation, and much closer to the
true value of $D$ than the value obtained by the length cutoff.

It should be noted that  the estimate
(\ref{e:d2}) only works if the sequence is
relatively smooth.  Cycle expansions with stability cutoff are not
particularly smooth as a function of the cutoff, for example see
Fig.~5 of ref.~\cite{DM97}.  This is because at each stage 
a small number of shadowing combinations are unbalanced
by the cutoff, and the number of such mismatches
varies rapidly with the cutoff.  In this case,
the convergence over the range $10^3$-$10^5$ is sufficiently smooth
to use (\ref{e:d2}), however a smaller spacing such as $2.5\times10^4$,
$5\times10^4$, $10^5$ is dominated by fluctuations.  

\section{Conclusion}

Our results indicate that cycle expansions may be used 
to calculate averages for intermittent systems with accuracy comparable
to direct simulations, as long as a stability cutoff is used.
Stability ordering has a great simplicity in that it requires
no knowledge of the dynamics, except what is contained in a finite cycle
set.  We conclude with a few possibilities for future directions.

First, it is good to see rapidly converging expansions, but another thing
to have rigorous limits to guarantee convergence.  Chaotic systems often
behave more nicely than it is possible to prove, however it would be
advantageous to extend the proofs of superexponential convergence for length
ordered cycle expansions of analytic hyperbolic systems to the stability
ordered case, hopefully allowing a wider class of dynamical systems.

The stability ordering exhibits imperfect shadowing, which can lead to scatter
in the results, as observed in Fig.~\ref{f:norm}.  One possible remedy
to this problem might be to replace the factor $1/\prod\Lambda$ by
$f(\prod\Lambda)/\prod\Lambda$ where the smoothing function $f$ moves
continuously from $1$ to $0$.  This must certainly improve the shadowing,
but it requires that cycles be found up to the largest stability at which
$f$ is non-zero, without utilizing these cycles fully.  For our diffusion
coefficient calculations, the results are smooth enough to use Aitken's
method, so additional smoothing is probably unnecessary.

We also note, that because the number of cycles less than a given $\Lambda$
is roughly $\Lambda/\ln\Lambda$, independent of the dimension of the space,
stability ordering should be applicable to high dimensional systems.  In
particular, the detailed structure of the dynamics need not be known, only
an algorithm for finding the cycles in the first place, for example tracing
out a long trajectory and looking for near repeats, which are then refined
by some form of Newton's method.  

Finally, it is not clear to what extent
the stability cutoff approach is applicable to quantum systems.  
It is not as easy to estimate the rate of convergence in this case, even
for nice hyperbolic flows, because the terms are complex, and the 
alternation of the Maslov phase
within families of cycles analogous to the $0^n+$ 
family studied above
crucial
for quantum convergence can lead to large errors in the stability
cutoff cycle expansion truncations~\cite{Sune97}.

\end{document}